\documentclass[10pt,twocolumn,british,showpacs]{revtex4}
\usepackage[latin1]{inputenc}
\usepackage{color}
\usepackage{graphicx}

\makeatletter

\providecommand{\tabularnewline}{\\}

\usepackage{color}
\usepackage{graphicx}
\usepackage{amssymb}

\usepackage[hypertex]{hyperref}
\newcommand{\App}{Appendix~}

\newcommand{\F}{Fig.~}

\usepackage{amsmath}
\usepackage{bm}
\renewcommand{\mathbf}{\bm}
\usepackage{ifthen}
\newboolean{includeNotes}
\setboolean{includeNotes}{false}

\usepackage{babel}

\usepackage{babel}
\makeatother
\begin{document}

\title{Nonequilibrium steady states in sheared binary fluids}

\author{P. Stansell$^{1}$}

\author{K. Stratford$^{2}$}

\author{J.-C. Desplat$^{3}$}

\author{R. Adhikari$^{1}$}

\author{M. E. Cates$^{1}$}

\affiliation{$^{1}$SUPA, School of Physics and $^{2}$EPCC, University of Edinburgh, JCMB Kings
Buildings, Mayfield Road, Edinburgh, EH9 3JZ, United Kingdom; $^{3}$Irish Centre
for High-End Computing, Dublin Institute for Advanced Studies, 5 Merrion Square,
Dublin 2, Ireland}

\begin{abstract}
We simulate by lattice Boltzmann the steady shearing of a binary fluid mixture undergoing
phase separation with full hydrodynamics in two dimensions. Contrary to some theoretical
scenarios, a dynamical steady state is attained with finite domain lengths $L_{x,y}$
in the directions ($x,y)$ of velocity and velocity gradient. Apparent scaling exponents
are estimated as $L_{x}\sim\dot{\gamma}^{-2/3}$ and $L_{y}\sim\dot{\gamma}^{-3/4}$.
We discuss the relative roles of diffusivity and hydrodynamics in attaining steady
state. 
\end{abstract}

\pacs{{\footnotesize 47.11.+j}}

\maketitle
Systems that are not in thermal equilibrium play a central role in modern statistical
physics, and arise in areas ranging from soap manufacture to subcellular biology
\cite{SFM}. Such systems include two important classes: those that are evolving
towards Boltzmann equilibrium (e.g., by phase separation following a temperature
quench), and those that are maintained in nonequilibrium by continuous driving (such
as a shear flow). Of fundamental interest, and surprising physical subtlety, are
systems combining both features --- such as a binary fluid undergoing phase separation
in the presence of shear. Here it is not known \cite{Onuki02,Cates99} whether coarsening
continues indefinitely, as it does without shear, or whether a steady state is reached,
in which the characteristic length scales $L_{x,y,z}$ of the fluid domain structure
attain finite $\dot{\gamma}$-dependent values at late times. (We define the mean
velocity as $u_{x}=\dot{\gamma}y$ so that $x,y,z$ are velocity, velocity gradient
and vorticity directions respectively; $\dot{\gamma}$ is the shear rate.) 

Experimentally, saturating length scales are reportedly reached after a period of
anisotropic domain growth \cite{Hashimoto888994,Onuki02}. However, the extreme elongation
of domains along the flow direction means that, even in experiments, finite size
effects could play an essential role in such saturation \cite{Bray03}. Theories
in which the velocity does not fluctuate, but does advect the diffusive fluctuations
of the concentration field, predict instead indefinite coarsening, with length scales
$L_{y,z}$ scaling as $\dot{\gamma}$-independent powers of the time $t$ since quench,
and (typically) $L_{x}\sim\dot{\gamma}tL_{y}$ \cite{Bray03}. In real fluids, however,
the velocity fluctuates strongly in nonlinear response to the advected concentration
field, and hydrodynamic scaling arguments, balancing either interfacial and viscous
or interfacial and inertial forces, predict saturation (e.g., $L\sim\dot{\gamma}^{-1}$
or $L\sim\dot{\gamma}^{-2/3}$) \cite{DoiM91,Onuki97,Cates99}. Given these experimental
and theoretical differences of opinion, computer simulations of sheared binary fluids,
with full hydrodynamics, are of major interest. 

The aforementioned scaling arguments cannot really distinguish one Cartesian direction
from another, but even in theories that can do so, a two dimensional (2D) representation,
suppressing $z$, is expected to capture the main physics \cite{Bray03}. (Without
shear, subtle non-scaling effects arise in 2D from the formation of disconnected
droplets \cite{WagnerYeomans98}, but shear seems to suppress these \cite{WagnerYeomans99}.)
Performing simulations in 2D is therefore a fair compromise, especially given the
extreme computational demands of the full 3D problem \cite{Cates99,CatesPT}. But,
apart from \cite{WagnerYeomans99,Gonnella01}, most numerical studies of binary fluids
under steady shear, even in 2D, neglect hydrodynamics altogether \cite{CorberiGL98,Lamura01b,Berthier01}.
Among fully hydrodynamic simulations (e.g., \cite{WagnerYeomans99,Gonnella01}),
only Wagner and Yeomans \cite{WagnerYeomans99} make a strong case for dynamical
steady states. In some cases these authors found complete remixing of the fluids
($L\rightarrow0$); in the remainder, finite size effects could not be excluded.
(To do so requires $L_{x,y}\ll\Lambda_{x,y}$ for a $\Lambda_{x}\times\Lambda_{y}$
simulation box.) The existence of nonequilibrium steady states, with finite $L_{x,y}$
in an infinite system, therefore remains an open question. 

In this letter we extend the hydrodynamic lattice Boltzmann (LB) studies of Refs.\cite{WagnerYeomans99,Gonnella01}
to much larger systems, which we then study over several decades of non-dimensionalized
shear rate. Unlike previous authors, we are able to give clear evidence of true dynamical
steady states, uncontaminated by finite size effects or other artifacts. (Finite
size effects typically result in quasi-laminar stripe domains which connect with
themselves after one or more circuits of the periodic boundary conditions \cite{WagnerYeomans99,CatesPT}.)
We then combine datasets using a quantitative scaling methodology developed for the
unsheared problem in \cite{Kendon01}; this allows scaling exponents to be estimated
using combined multi-decade fits. By this method we find apparent scaling exponents
$L_{x}\sim\dot{\gamma}^{-2/3},L_{y}\sim\dot{\gamma}^{-3/4}$, sustained over six
decades of shear rate. 

Our basic LB algorithm for binary fluids is essentially as reported in \cite{Kendon01}
(see also \cite{Swift96}) on a D2Q9 lattice. Additionally we exploit recent algorithmic
advances \cite{WagPag,Adhikari05} that overcome the intrinsic fluid velocity limit
of LB by using blockwise translating lattice slabs connected by multiple Lees-Edwards
boundary conditions \cite{WagPag}. (Details of our boundary conditions, with validation
data, appear in \cite{Adhikari05}.) One technical problem that remains within our
LB scheme concerns the role of order parameter diffusivity. In the hydrodynamic coarsening
regimes of main interest (late times, modest shear rates) this diffusivity should
always maintain local equilibrium across fluid interfaces, but never transport significant
material across the interior of domains \cite{Kendon01}. Under shear when domains
are extremely anisotropic, compromise becomes inevitable. We discuss below the implications
of this for the interpretation of our apparent scaling exponents. 

\begin{figure}
\begin{center}\includegraphics[%
  width=1.0\columnwidth]{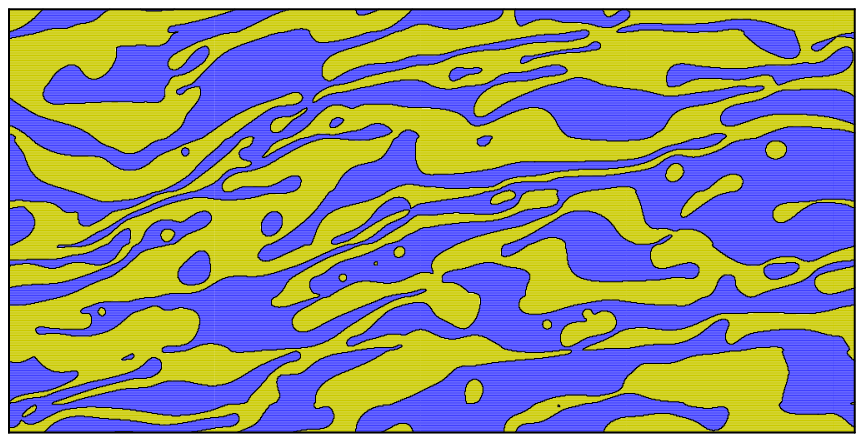}\end{center}

\begin{center}\includegraphics[%
  width=1.0\columnwidth]{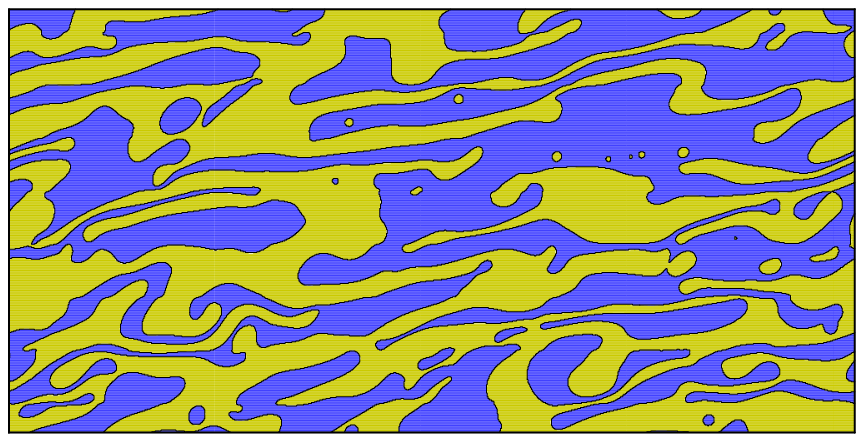}\end{center}

\caption{\label{fig:PhiFields}Snapshots of the steady-state order-parameter for R020 with
$\dot{\gamma}=0.0005$ at dimensionless times $t\dot{\gamma}=60$ and $255$ (upper
and lower).}

\ifthenelse{\boolean{includeNotes}}{

\textcolor{green}{Plots provided by JC and Kevin.}

}

\end{figure}

The governing equations that our LB scheme approximates are the Cahn-Hilliard equation
for the compositional order parameter $\varphi$, and the incompressible ($\nabla_{\alpha}u_{\alpha}=0$)
Navier-Stokes equation for the velocity $u_{\alpha}$ in an isothermal fluid of unit
mass density: \begin{eqnarray}
\left(\partial_{t}u_{\alpha}+u_{\beta}\nabla_{\beta}u_{\alpha}\right)+\nabla_{\alpha}p-\nu\nabla^{2}u_{\alpha}-\phi\nabla_{\alpha}\mu & = & 0\label{two}\\
\partial_{t}\varphi+\nabla_{\alpha}\left(\varphi u_{\alpha}-M\nabla_{\alpha}\mu\right) & = & 0\label{three}\end{eqnarray}
 Here $p$ is pressure (related in LB to density fluctuations, which remain small,
by an ideal gas equation of state \cite{Kendon01}); $\nu$ is the kinematic viscosity;
$M$ is the order-parameter mobility, and $\mu=B\varphi\left(\varphi^{2}-1\right)-\kappa\nabla^{2}\varphi$,
the chemical potential. $B$ and $\kappa$ are positive constants; the interfacial
tension is $\sigma=(8\kappa B/9)^{1/2}$ and the interfacial width is $\xi_{0}=(2\kappa/B)^{1/2}$
\cite{Kendon01}. LB control parameters are $B$, $\kappa$, $M$ and $\nu$ alongside
the steady shear rate $\dot{\gamma}$. 

Blockwise sheared boundary conditions are imposed \cite{Adhikari05} such that $\int_{0}^{\Lambda_{y}}\nabla u_{x}\, dy=\Lambda_{y}\dot{\gamma}$.
A fluctuating local velocity field can then arise by nonlinearity, as in experiments
\cite{WagPag}. (We neglect \emph{thermal} fluctuations in our fluid, as appropriate
for dynamics near a zero-temperature fixed point \cite{Bray94}.) Under shear we
define length scales $L_{x,y}$ using a gradient statistic for $\varphi$ that measures
the mean distance between interfaces lying across to the chosen direction \cite{WagnerYeomans99}.
We also define $L_{\Vert,\bot}$, with $L_{\Vert}>L_{\bot}$ by reference to appropriate
principal axes. (Ref. \cite{Berthier01} finds universality, in a related but nonhydrodynamic
system, when lengths are scaled with $L_{\Vert,\bot}$ but not with $L_{x,y}$.) 

The physics of binary fluid demixing, with no shear and low enough diffusivity $M$,
can be nondimensionalized via a single length scale $L_{0}=\nu^{2}/\sigma$ and time
scale $T_{0}=\nu^{3}/\sigma^{2}$. These are, up to dimensionless prefactors \cite{Kendon01},
the domain size and time after quench at which an interfacial/viscous balance in
the coarsening dynamics (viscous hydrodynamic regime, VH) crosses over to an interfacial/inertial
balance (inertial hydrodynamic, IH). By scaling the domain length $L$ and times
$t$ by $L_{0}$ and $T_{0}$, multiple datasets were shown to merge, giving a universal
crossover between VH and IH regimes \cite{Kendon01,Pagonabarraga02}. 

Accordingly, in our search for nonequilibrium steady states, we nondimensionalize
the shear rate as $\dot{\gamma}T_{0}$ and domain sizes as $L_{x,y,\Vert,\bot}/L_{0}$.
One can then expect plots of length against strain rate (or its inverse, as used
below) to show data collapse whenever diffusivity is small. One possibility \cite{Cates99}
is that all such plots might exhibit the same crossover from VH to IH behaviour as
would be found by substituting $t=\dot{\gamma}^{-1}$ in the universal scaling plot
of $L/L_{0}$ against $t/T_{0}$. This would give $L_{x,y,\Vert,\bot}\sim L$ with
$L/L_{0}\sim(\dot{\gamma}T_{0})^{-1}$ at large shear rates and $L/L_{0}\sim(\dot{\gamma}T_{0})^{-2/3}$
at small ones. Alternatively, some single power law could persist at all $\dot{\gamma}T_{0}$
\cite{DoiM91,Cates99}; and/or there could be different exponents in the different
directions; or there might be no steady state at all \cite{Bray03}. 

\if{
\section{Simulation}

\subsection{The LBM algorithm}

The governing equations were solved numerically in two dimensions using the lattice
Boltzmann method (LBM) on a D2Q9 lattice for both the fluid and order-parameter.
The equation of state is \(p=c_{s}^{2}\rho\), where \(c_{s}\) is the speed of sound.
In the low Mach number limit solutions using the LBM approximate those of the compressible
Navier-Stokes equations. In the LBM discrete fluid and order-parameter distributions,
denoted by \(f_{i}\) and \(g_{i}\) respectively, are updated by the following iterative
scheme. The macroscopic variables are calculated\[
\left[\begin{array}{c}
\rho\\
\phi\\
\rho u_{\alpha}\\
S_{\alpha\beta}\\
j_{\alpha}\\
\Phi_{\alpha\beta}\end{array}\right]=\sum_{i=0}^{8}\left[\begin{array}{c}
f_{i}\\
g_{i}\\
c_{i\alpha}f_{i}\\
Q_{i\alpha\beta}f_{i}\\
c_{i\alpha}g_{i}\\
Q_{i\alpha\beta}g_{i}\end{array}\right],\]
where \(j_{\alpha}\) is the order-parameter flux and \(Q_{i\alpha\beta}=c_{i\alpha}c_{i\beta}-c_{s}^{2}\delta_{\alpha\beta}\).
The equilibrium values for the relaxed variables are calculated by\[
\left[\begin{array}{c}
S_{\alpha\beta}^{(0)}\\
j_{\alpha}^{(0)}\\
\Phi_{\alpha\beta}^{(0)}\end{array}\right]=\left[\begin{array}{c}
\rho u_{\alpha}u_{\beta}+P_{\alpha\beta}\\
\phi u_{\alpha}\\
2M\mu\delta_{\alpha\beta}/\rho_{0}+\phi u_{\alpha}u_{\beta}\end{array}\right],\]
where the thermodynamic force is modelled by \(F_{\alpha}=\partial_{\alpha}P_{\alpha\beta}\),
with \(P_{\alpha\beta}=\left(\frac{1}{2}B\phi^{2}+\frac{3}{4}B\phi^{4}-\kappa\phi\nabla^{2}\phi-\frac{1}{2}\kappa\left(\nabla\phi\right)^{2}\right)\delta_{\alpha\beta}+\kappa\left(\partial_{\alpha}\phi\right)\left(\partial_{\beta}\phi\right)\)
as implied by the definition of \(\mu\). Next, the relaxation toward equilibrium values
at the appropriate rates is carried out according by\[
\left[\begin{array}{c}
S_{\alpha\beta}^{+}\\
j_{\alpha}^{+}\\
\Phi_{\alpha\beta}^{+}\end{array}\right]=\left[\begin{array}{l}
S_{\alpha\beta}-\tau^{-1}\left(S_{\alpha\beta}-S_{\alpha\beta}^{(0)}\right)\\
j_{\alpha}^{(0)}\\
\Phi_{\alpha\beta}^{(0)}\end{array}\right]\]
where the relaxation time for the fluid is \(\tau=\nu/c_{s}^{2}\Delta t+1/2\) and
that for the order-parameter is set to unity. The post collision distributions are
calculated by\[
\left[\begin{array}{c}
f_{i}^{+}\\
g_{i}^{+}\end{array}\right]=w_{i}\left(\left[\begin{array}{c}
\rho\\
\phi\end{array}\right]+\left[\begin{array}{c}
\rho u_{\alpha}\\
j_{\alpha}^{+}\end{array}\right]\frac{c_{i\alpha}}{c_{s}^{2}}+\left[\begin{array}{c}
S_{\alpha\beta}^{+}\\
\Phi_{\alpha\beta}^{+}\end{array}\right]\frac{Q_{i\alpha\beta}}{2c_{s}^{4}}\right),\]
and finally the post collision distributions are streamed to their \(t+\Delta t\)
positions\[
\left[\begin{array}{c}
f_{i}\\
g_{i}\end{array}\right](\mathbf{x}+\mathbf{c}_{i},t+\Delta t)=\left[\begin{array}{c}
f_{i}^{+}\\
g_{i}^{+}\end{array}\right](\mathbf{x},t).\]

It is recognised that this scheme gives rise to erroneous terms in the simulated
equations. They are caused by the coupling of the order-parameter to the fluid through
the stress tensor \(P_{\alpha\beta}\), and by choosing to relax the order-parameter
distributions through \(M\) directly in \(\Phi_{\alpha\beta}^{(0)}\) instead of relaxing
both \(j_{\alpha}\) and \(\Phi_{\alpha\beta}\). For a discussion of these errors and
their relative magnitudes see \cite{Kendon01}. No erroneous terms are present in
the simulated equations if the coupling is performed directly through the body-force
density \(F_{\alpha}=\rho_{0}X_{\alpha}\) and both \(j_{\alpha}\) and \(\Phi_{\alpha\beta}\)
are relaxed as described in outline in \App\ref{App:forceCoupling}.

The algorithm used to implement the externally imposed uniform shear within the LBM
scheme was that described by \cite{Adhikari05} which allows cyclic boundary conditions
to be applied at all edges of the computational domain.
} \fi

All simulations reported here were done for fully symmetric quenches on a $\Lambda_{x}\times\Lambda_{y}=1024\times512$
lattice, with up to $t=6\times10^{5}$ updates. Parameters $\xi_{0},\sigma,M,\nu$
(Table~\ref{tab:dataSets}) and $\dot{\gamma}$ were chosen, following \cite{Kendon01},
so that: interfaces are wide enough to be resolved (restrictions on $\xi_{0}$);
fluid flow is slow enough for advected interfaces to be in local equilibrium (restrictions
on $\sigma$ and $\dot{\gamma}$); the diffusivity is low enough not to contaminate
steady-state length scales, e.g., detectable as a strong residual $M$ dependence
(restriction on $M$). Also, these steady-state lengths must be sufficiently small
to avoid finite size effects (quantified below) and the code must run stably for
long enough to achieve steady state, typically $t\geq10^{5}$ updates. 

Acceptable shear rates were found to be $1.25\times10^{-4}\leq\dot{\gamma}\leq2\times10^{-3}$
(in lattice units). Higher values gave inaccuracies as listed above; lower values
gave unacceptably long run times. As in the unsheared case \cite{Kendon01} judicious
combinations of $\xi_{0}$, $\sigma$, $M$ and $\nu$ allow systems spanning several
decades in $L/L_{0}$ and $\dot{\gamma}T_{0}$ to be accurately studied, by exploiting
LB's ability to vary $L_{0}$ and $T_{0}$ alongside $\dot{\gamma}$. 

\begin{table}
\begin{center}\begin{tabular}{ccccccc}
\hline 
Name&
 $\nu$&
 $M$&
 $\sigma_{\textrm{theory}}$&
 $\sigma_{\textrm{meas}}$&
 $L_{0}$&
 $T_{0}$\tabularnewline
\hline
R028&
 1.41&
 0.05&
 0.063&
 0.055&
 36.1&
 927\tabularnewline
R022&
 0.5&
 0.25&
 0.047&
 0.042&
 5.95&
 70.9\tabularnewline
R029&
 0.2&
 0.15&
 0.047&
 0.042&
 0.952&
 4.54\tabularnewline
R020&
 0.025&
 2&
 0.0047&
 0.0042&
 0.149&
 0.886\tabularnewline
R030&
 0.00625 &
 1.25&
 0.0047&
 0.0042&
 0.00930&
 0.0138\tabularnewline
R019&
 0.0014&
 4&
 0.0024&
 0.0021&
 0.000933&
 0.000622\tabularnewline
R032&
 0.0005&
 5&
 0.00094&
 0.00083&
 0.000301&
 0.000181 \tabularnewline
\hline
\end{tabular}\end{center}

\caption{\label{tab:dataSets}Parameter sets used in simulations, along with $L_{0}$ and
$T_{0}$. In all cases $\xi_{0}=1.13$. (See \cite{Kendon01} for discussion on the
difference between the theoretical and measured values of $\sigma$; the measured
ones are used to determine $L_{0},T_{0}$).}

\ifthenelse{\boolean{includeNotes}}{

\textcolor{green}{Data produced by \textasciitilde{}/physics/\-Edinburgh/\-binaryMixtures/\-notes/\-Fortran/\-crossOver.f
and cut'n'paste from \textasciitilde{}/physics/\-Edinburgh/\-binaryMixtures/\-notes/\-binaryMixturesII.lyx.}

}

\end{table}

\begin{figure}
\begin{center}\includegraphics[%
  width=0.90\columnwidth]{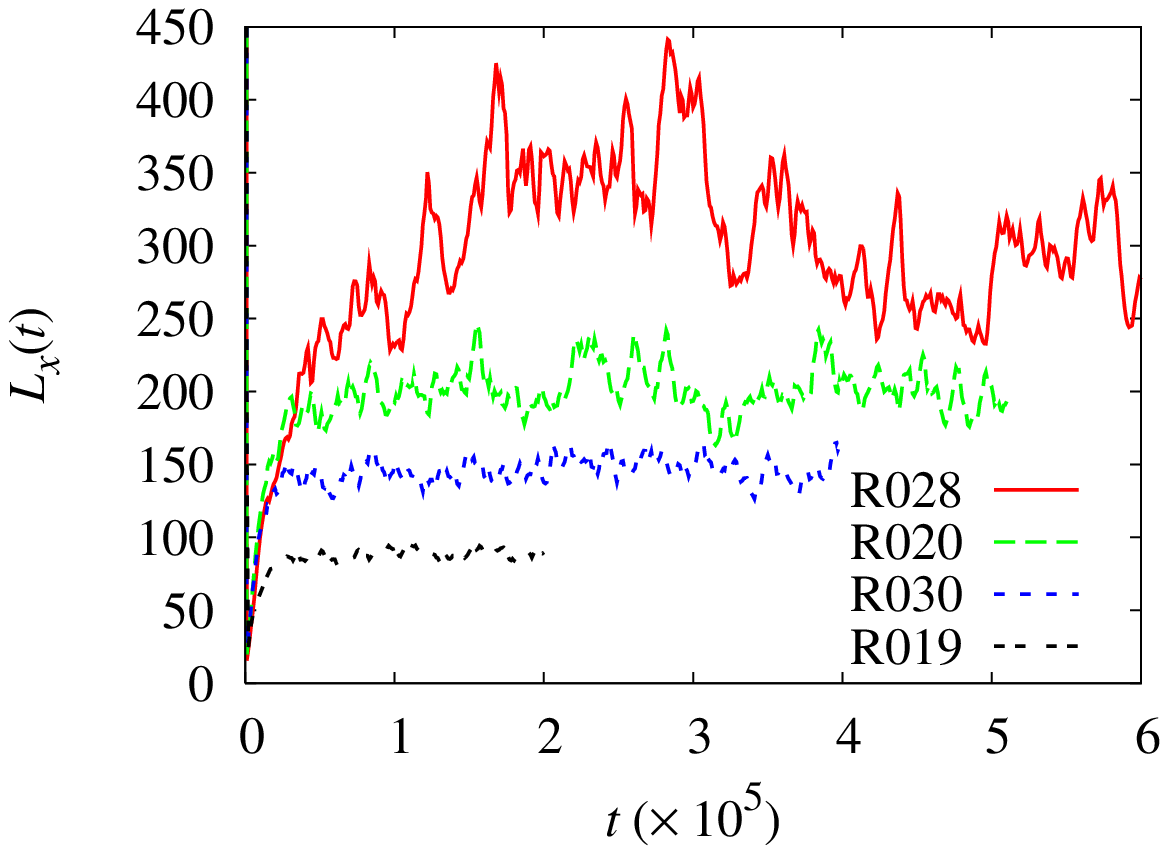}\end{center}

\begin{center}\includegraphics[%
  width=0.90\columnwidth]{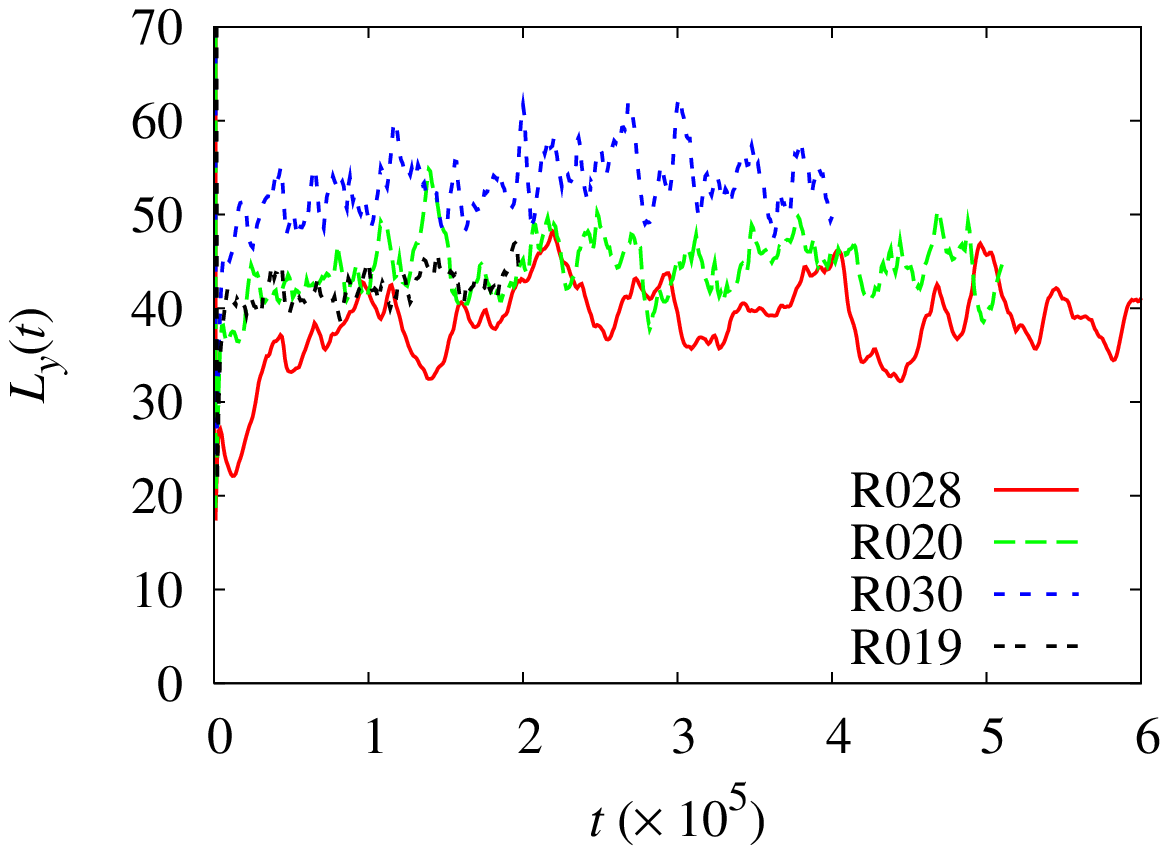}\end{center}

\caption{\label{fig:timeSeries}Plots of $L_{x}(t)$ and $L_{y}(t)$ in lattice units, for
various runs with $\dot{\gamma}=5\times10^{-4}$.}

\ifthenelse{\boolean{includeNotes}}{

\textcolor{green}{Plot produced by \textasciitilde{}/papers/\-unfinished/\-spinDecomShear2D/\-gnuplot/\-plotTimeSeries.gp.}

}

\end{figure}

\F\ref{fig:PhiFields} shows two snapshots of the order-parameter field for R020
with $\dot{\gamma}=5\times10^{-4}$ after a steady state had been reached. The snapshots
are at dimensionless times $t\gamma=60$ and $t\gamma=255$, for the upper and lower
plots respectively. \F\ref{fig:timeSeries} shows unscaled time-series for $L_{x}$
and $L_{y}$ from a representative set of simulations with $\dot{\gamma}=5\times10^{-4}$.
Both figures show decisive evidence of length-scale saturation in a regime that seems
safely clear of any finite size effects. A number of tests were performed in which
all run parameters were held constant except the lattice dimensions which were changed
in the ranges $\Lambda_{x}=512$ to $2048$ and $\Lambda_{y}=256$ to $1024$. From
the results of these tests we conclude that finite-size effects are fully under control
when $L_{x,y}\leq\Lambda_{x,y}/4$, a criterion extensively benchmarked in unsheared
systems \cite{Kendon01}. At the same time, $L_{x,y}\geq30$ in lattice units, well
clear of discretisation artifacts. However, the thin fluid threads visible in \F\ref{fig:PhiFields}
mean that residual diffusion cannot entirely be ruled out; we return to this point
below. 

\begin{figure}
\begin{center}\includegraphics[%
  width=0.40\textwidth]{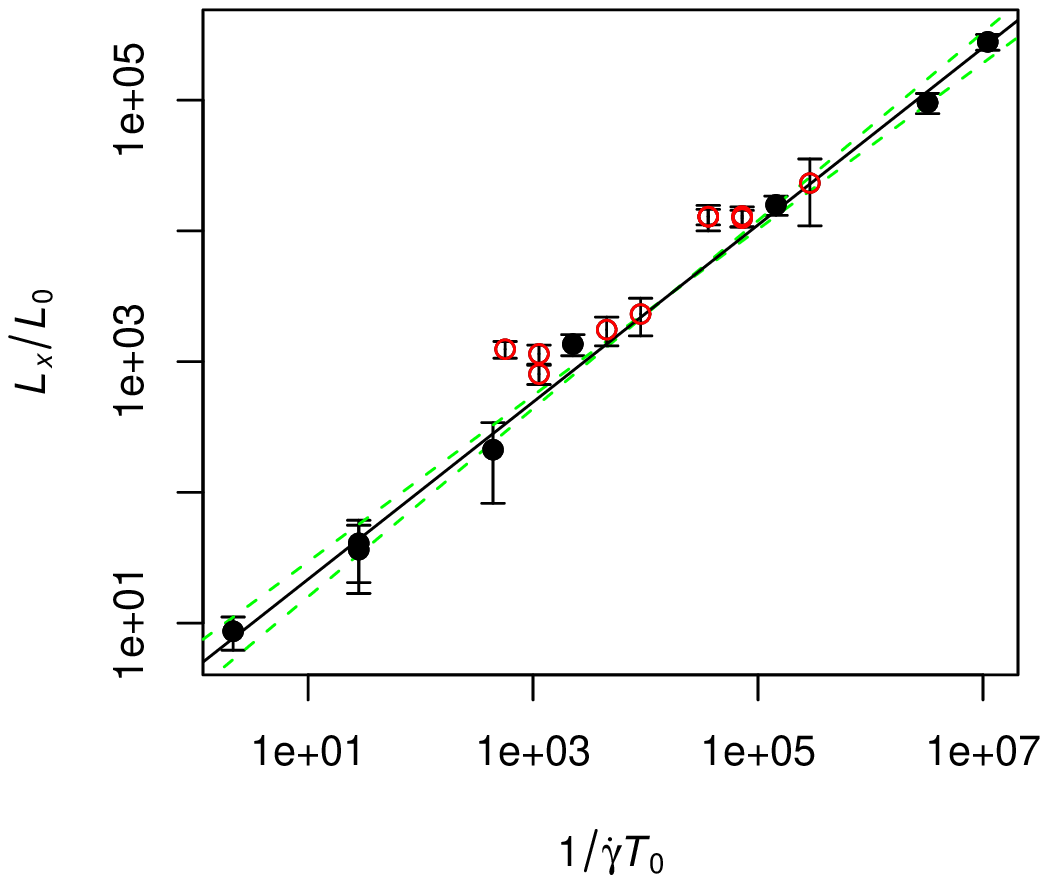}\end{center}

\begin{center}\includegraphics[%
  width=0.40\textwidth]{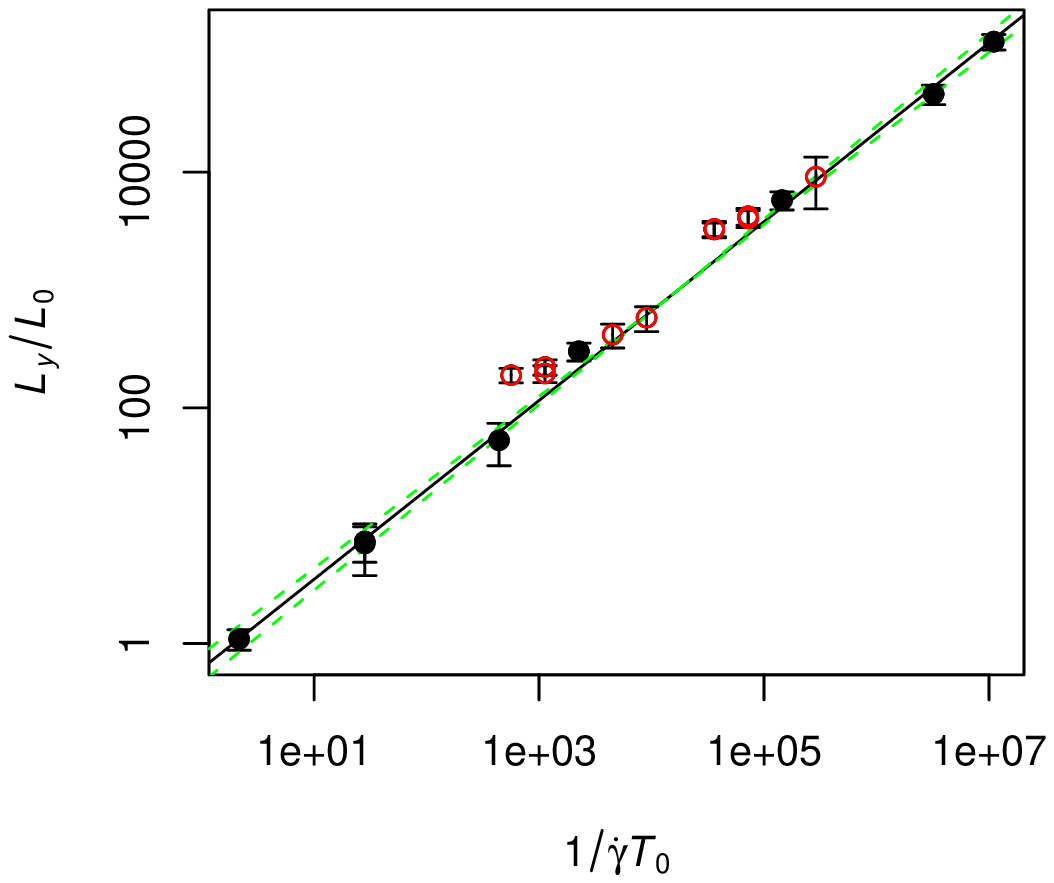}\vspace{-1.5em}\end{center}

\caption{\label{fig:scalingsVI}Dimensionless scaling plots of lengths vs shear rate. Error
bars as shown give the 95\% confidence limits (dashed lines) on the fitted linear
regression (solid line). Solid symbols: $\dot{\gamma}=5\times10^{-4}$. Open symbols
have $\dot{\gamma}$ one of, from left to right, $2\times10^{-3}$, $10^{-2}$, $2.5\times10^{-4}$
or $1.25\times10^{-4}$. The fit used only the data for $\dot{\gamma}=5\times10^{-4}$.
Points with two symbols at the same $1/\dot{\gamma}T_{0}$ denote variations in the
diffusivity $M$.}

\ifthenelse{\boolean{includeNotes}}{

\textcolor{green}{Plot produced by \textasciitilde{}/physics/\-Edinburgh/\-binaryMixtures/\-bluegeneData/\-2D/\-R-analysis/\-scalingL.R.}

}

\end{figure}

\F\ref{fig:scalingsVI} shows dimensionless scaling plots of steady-state length
scales $L_{x,y}$ against shear rate, for a series of runs in which $\dot{\gamma}=5\times10^{-4}$
with variable $L_{0},T_{0}$ (solid symbols). $L_{x,y}$ were obtained as the temporal
means of the time-series data $L_{x}(t)$ and $L_{y}(t)$, after discarding data
for which $t<10^{5}$. Uncertainties in $L_{x,y}$ (also $L_{\Vert,\bot}$) were
found using 999 bootstrap replicas of the time-series data. These uncertainties,
which provide the relative error bars indicated in the scaling plots, were then used
to weight the individual data-points in a linear least-squares regression from which
estimates were obtained of the intercept and gradient of the straight lines shown
on the log-log plots. Fitted exponents for $L_{x}$ and $L_{x}$ are $-0.678\pm0.042$
and $-0.759\pm0.029$; and for $L_{\Vert}$ and $L_{\bot}$, $-0.678\pm0.039$ and
$-0.756\pm0.030$ (data not shown) \cite{footint}. The $95$\% confidence level
admits the appealing ansatz of fractional power laws $L_{x}\sim L_{\Vert}\sim\dot{\gamma}^{-2/3}$
and $L_{y}\sim L_{\bot}\sim\dot{\gamma}^{-3/4}$. 

\if{
\(L_{x}/L_{0}\sim L_{\Vert}/L_{0} \sim \left(\dot\gamma T_{0}\right)^{-2/3}\) and \(
L_{y}/L_{0}\sim L_{\bot}/L_{0} \sim \left(\dot\gamma T_{0}\right)^{-3/4}\).
}\fi
However, caution is warranted in presenting these apparently clean scaling laws.
Firstly, were the dynamic length scales to be found by substituting $t=\dot{\gamma}^{-1}$
in the shear-free coarsening plot as suggested above \cite{Cates99}, then a slow
crossover from VH ($L\sim\dot{\gamma}^{-1}$) to IH ($L\sim\dot{\gamma}^{-2/3}$)
would affect this entire range of $\dot{\gamma}T_{0}$ \cite{Kendon01}. Fitting
$L_{x,y}$ data to simple power laws might therefore be misleading. Secondly, we
also show in \F\ref{fig:scalingsVI} two datasets found by varying $\dot{\gamma}$
with other parameters fixed. For both $L_{x}$ and $L_{y}$, if taken in isolation
these sets would suggest smaller slopes than seen for the main plot. Such deviations
were found previously in the unsheared case \cite{Kendon01}, and argued to be a
signature of residual diffusion, with each dataset asymptoting onto the global trend
line from above left. 

The upward curvature of these two datasets, and the fact that they appear to asymptote
onto (rather than cross) the best fit line found for $\dot{\gamma}=5\times10^{-4}$,
offers some reassurance, but no guarantee, that the latter is uncontaminated by residual
diffusion at the domain scale. If so, our stationary states stem not from diffusion
(although that would itself be interesting) but from the hydrodynamic balance between
the stretching, breaking and coalescence of domains. Further evidence for this comes
from study of the evolving $\varphi$ field, in which all of these effects are visible
but (with our parameters) large-scale diffusion is not. Also, all the simulations
reported here quench at $t=0$ from a noisy but uniform state. The system then passes
through, and apparently leaves, a diffusive regime prior to the hydrodynamic one
that gets cut off by the presence of shear. As a final check, we show directly the
effect of varying $M$ in two runs shown in \F\ref{fig:scalingsVI} (where more
than one symbol occurs at the same $\dot{\gamma}T_{0}$). The lower data-points were
found by reducing $M$ by a factor 2 or more from nearby runs. The resulting shifts
are modest, although not entirely negligible --- particularly towards the bottom
left of the plots. Thus it remains possible that further reduction in $M$ (not practical
numerically at present) could reveal a significant kink on these plots, as might
be expected near a VH to IH crossover. 

In conclusion, although the apparent scaling exponents reported above are interesting
and merit both theoretical investigation and experimental tests, the main significance
of our work is in the unambiguous demonstration of nonequilibrium steady states in
sheared binary fluids. Since theories that neglect velocity fluctuations do not predict
such states \cite{Bray03}, hydrodynamics appears to play an essential role. (This
remains true even if diffusion is not negligible as considered above.) A key question
is whether such steady states persist in three dimensions. Although the physics of
stretching, breakup and coalescence is captured in 2D, in 3D there can remain tubular
connections between domains in the vorticity direction; these could remain relatively
unaffected by shear. This might leave open a route to continuous coarsening that
is topologically absent in two dimensions. We hope to address the 3D case in future
simulations. 

\emph{Acknowledgements:} We thank Ignacio Pagonabarraga and Alexander Wagner for
discussions. Work funded by EPSRC GR/S10377 and GR/R67699 (RealityGrid). One of us
(JCD) would like to acknowledge the Irish Centre for High-End Computing for access
to their computing facilities (SFI grant 04/HEC/I584s1) and support.


\begin{thebibliography}{10}
\bibitem{footint}The fitted values for the \emph{intercepts} on the log-log scaling plots are: $0.657\pm0.176$
for $L_{x}$, $-0.214\pm0.1216$ for $L_{y}$, $0.742\pm0.166$ for $L_{\Vert}$,
and $-0.214\pm0.125$ for $L_{\bot}$, where the estimated errors represent 95\%
confidence intervals.
\end{thebibliography}
\end{document}